\documentclass[10p
 t,conf]{IEEEtran}

 \usepackage[left=0.625in,right=0.625in,top=0.75in,bottom=0.99in]{geometry}
\usepackage{cite}
\usepackage{amsmath,amssymb,amsfonts}
\usepackage{graphicx}
\usepackage{color}
\usepackage{subfig}
\usepackage{stackengine}
\usepackage{comment}
\usepackage[normalem]{ulem}
\usepackage{tabularx}
\usepackage{bm}
\usepackage{dsfont}
\usepackage{mathtools}
\usepackage[font=scriptsize]{caption}

\usepackage[singlespacing]{setspace} 
\usepackage[colorlinks, citecolor=blue, bookmarks=false]{hyperref} 

\usepackage[flushleft]{threeparttable} 
\usepackage{booktabs}

\usepackage{textcomp}
\usepackage{xcolor}

\usepackage[algo2e]{algorithm2e} 
\SetKwRepeat{Do}{do}{while}%

\SetCommentSty{mycommfont}
\usepackage{algorithm}
\usepackage{algpseudocode}

\def\BibTeX{{\rm B\kern-.05em{\sc i\kern-.025em b}\kern-.08em
    T\kern-.1667em\lower.7ex\hbox{E}\kern-.125emX}}
    
\algnewcommand\algorithmicforeach{\textbf{for each}}
\algdef{S}[FOR]{ForEach}[1]{\algorithmicforeach\ #1\ \algorithmicdo}

\usepackage{orcidlink}

\usepackage{xspace}
\usepackage{ifthen}
\usepackage[inline]{enumitem}

\usepackage[utf8]{inputenc}

\usepackage{algcompatible}
\usepackage{algorithmicx}

\newboolean{showcomments}
\setboolean{showcomments}{true}
\ifthenelse{\boolean{showcomments}}
{
}

\usepackage{subfig}
\usepackage{fancyhdr}

\begin{document}

\renewcommand{\algorithmiccomment}[1]{\textit{#1}}

\title{\huge A New Realistic Platform for Benchmarking and Performance Evaluation of DRL-Driven and Reconfigurable SFC Provisioning Solutions}

\author{
\IEEEauthorblockN{Murat Arda Onsu$^1$, Poonam Lohan$^1$, Burak Kantarci$^1$, Emil Janulewicz$^2$, Sergio Slobodrian$^2$,}\\
\IEEEauthorblockA{\textit{$^1$University of Ottawa, Ottawa, ON, Canada}\\
\textit{$^2$Ciena, 383 Terry Fox Dr,
Kanata, ON K2K 2P5, Canada}\\
$^1$\{monsu022, ppoonam, burak.kantarci\}@uottawa.ca,~$^2$\{ejanulew,sslobodr\}@ciena.com}
\vspace{-0.3in}
}

           



\maketitle
\begin{abstract}

 Service Function Chain (SFC) provisioning stands as a pivotal technology in the realm of 5G and future networks. Its essence lies in orchestrating VNFs (Virtual Network Functions) in a specified sequence for different types of SFC requests. Efficient SFC provisioning requires fast, reliable, and automatic VNFs' placements, especially in a network where massive amounts of SFC requests are generated having ultra-reliable and low latency communication (URLLC) requirements.  
Although much research has been done in this area, including Artificial Intelligence (AI)  and Machine Learning (ML)-based solutions, this work presents an advanced Deep Reinforcement Learning (DRL)-based simulation model for SFC provisioning that illustrates a realistic environment. The proposed simulation platform can handle massive heterogeneous SFC requests having different characteristics in terms of VNFs chain, bandwidth, and latency constraints. Also, the model is flexible to apply to networks having different configurations in terms of the number of data centers (DCs), logical connections among DCs, and service demands. The simulation model components and the workflow of processing VNFs in the SFC requests are described in detail. Numerical results demonstrate that using this simulation setup and proposed algorithm, a realistic SFC provisioning can be achieved with an optimal SFC acceptance ratio while minimizing the E2E latency and resource consumption. 


\end{abstract}


\section{Introduction} \label{sec:1}


Service Function Chaining (SFC) utilizes NFV benefits \cite{new_4}, orchestrating a sequence of Virtual Network Functions (VNFs) for diverse applications like Cloud Gaming (CG), Augmented Reality (AugR), Video Streaming (VS), VoIP, Massive IoT (MIoT), and Industry 4.0 (Ind4.0) \cite{13}\cite{2}. As challenges persist, including resource allocation optimization, sequential VNF processing, scalability, and  End-to-End (E2E) delay constraints, several researchers have proposed advanced algorithms, such as deep learning (DL) and reinforcement learning, for optimal and automated SFC provisioning \cite{rel1}. 
Studying algorithm impacts and proposing new methodologies demands reliable simulation tools. Platforms such as NS3 \cite{NS3} are utilized for packet flow analysis, rather than VNF function placements. In these simulation models, VNF functions are inherent system attributes, assumed installed upon hardware creation. Yet, for SFC provisioning, detailed consideration of resource allocation, latency, function chain order, and attributes is vital for request management. Therefore, this study proposes a novel simulation platform that handles the SFC provisioning task using VNF function placement while considering proper allocation in sequence, resource availability, and latency constraints. 


This platform includes various VNFs, SFCs, data centers (DCs), and logical connections among DCs. SFC provisioning tasks need to account for VNF processing and transmission (TX). VNFs processing must follow the SFC sequence order. VNF installation depends on DC resource availability, while processing duration is determined by VNF characteristics. Packet TX happens concurrently with VNFs processing across different DCs if VNFs in SFC are allocated to different DCs and to the destination after all VNFs have been processed. 

To ensure efficient SFC provisioning, an effective VNF placement algorithm must address challenges including sequential VNF execution, resource efficiency, accommodating high demands, and meeting E2E delay requirements \cite{iee4}\cite{10}. 
The proposed platform features a flexible simulation model that operates independently of placement or provisioning algorithms, allowing for the seamless integration of any AI model while preserving the integrity of simulation runtime. The Deep Reinforcement Learning (DRL) model, known for its adept decision-making and adaptability within complex systems, is employed in the simulation \cite{finalR1}. It utilizes a DL architecture to make decisions and learn through trial and error exploration of the environment \cite{DRL}. To strengthen the algorithm's robustness, priority points are allocated to VNF functions based on their remaining E2E latency, chain order, and other SFC states. Decisions are guided by both the outputs of a DRL model and the priority points. Given the DRL model's reliance on exploration and environmental observation, an advanced architecture featuring multiple input layers for comprehensive observation is introduced. Throughout simulations,
data on E2E delay and SFC acceptance ratio  are collected,
demonstrating superior performance of the proposed DRL
model over the benchmark in both metrics. The main contributions of this work are as follows:

\begin{itemize}
    \item Introduce a novel platform for simulating and benchmarking DRL-based SFC provisioning, facilitating comprehensive data collection during runtime, covering SFCs, VNFs, bandwidth, and DCs. Its robust components and workflow ensure dependable results, making it a valuable resource for exploring various methodologies and models in SFC provisioning research.
    \item An advanced DRL architecture is proposed for SFC provisioning. This architecture incorporates various input states, enabling the model to expand its observations, thus enhancing its decision-making capabilities.
    \item DCs receive priority scores determined by their resources and incoming SFC demands, while VNFs get scores based on attributes, position in the chain, and remaining E2E delay. The DRL model employs these scores to optimize action selection.
\end{itemize}

The paper is structured as follows: Section II reviews related works. Section III presents the system model and problem formulation. Section IV outlines the simulation model. Section V introduces the VNF placement using DRL with priority points for SFC provisioning. Section VI presents the results, with conclusions in Section VII. 

\section{Related Works} \label{sec:rel}

In research \cite{glob_rel_1}, a novel placement emulator framework (PEF), integrated with a Placement Abstraction Layer (PAL),  is introduced as a tool that enables the execution of calculated placements using genuine, container-based VNFs within real-world network topologies. On the other hand, researchers in \cite{glob_rel_3} study novel management architecture for 5G service-based core networks based on NFV and SDN. The framework includes features like dynamic service migration, logical dedicated networks, and flexible orchestration of network functions, aiming to support a service-centric network in 5G era. Furthermore, in research \cite{glob_rel_4}, A novel model based on capability profiles is proposed for SFCs within SDN/NFV architecture by studying performance evaluation challenges and offering adaptability for infrastructure implementations and service capacity scaling.

The study in \cite{glob_rel_5} delves into offering SFC as a Service (SFCaaS) in NFV setups. It presents an Integer Linear Program and two heuristics to optimize resource allocation, cost reduction, and profit maximization. The study highlights key factors like instance costs, VNF performance, and cost-effectiveness of smaller instances. It develops a C-based simulator for physical topology simulation and SFC request translation and mapping. Further, the research \cite{glob_rel_7} introduces SoftIoT, an energy-efficient SDN/NFV-based IoT network surpassing current solutions in energy consumption, packet delivery, network lifetime, and overhead. Further research is needed on real-testbed implementation and optimized designs for interference-prone wireless networks to address IoT network resource and energy constraints.

In \cite{ns_rel_1}, researchers discuss promising aspects of 5G networks, the need for a comprehensive simulation environment, and contributions related to evaluating radio access technologies and SDN impact on the transport segment. Moreover, a case study of using ns-3 to realize various architectures for Data Center Networks (DCNs) and study their performance is presented in \cite{ns_rel_2}. This study primarily evaluates routing algorithms in DCNs using NS3. It focuses on memory usage, CPU utilization, and data transfer delay. Additionally, OMNET++-based simulation tool designed for applications utilizing vehicular and edge-cloud concepts is explored in \cite{ns_rel_4}. 

Numerous studies address simulation platforms, highlighting limitations. In \cite{glob_rel_1}, PAL and PEF are discussed, which permit diverse algorithm and backend integration; however, emulation efficacy relies on precise network model and VNF accuracy. \cite{glob_rel_3} underscores latency reduction goals but notes that dynamic functions and resource allocation might introduce latency. The work in \cite{glob_rel_4} relies on analytical modeling and simulations to evaluate the proposed model's effectiveness causing the lack of real-world testing. The study in, \cite{ns_rel_1} offers adaptable 5G simulation for analyzing varied network configurations and their performance outcomes. While useful, ns3's capacity for simulating real-world behaviors may be limited in certain contexts, particularly regarding research on VNF placement on hardware components. While DCNs are also studied with the ns3 simulator \cite{ns_rel_2}, there exist accuracy limitations in simulating certain real-world network behaviors. 


\section{System Model and Problem Formulation} \label{sec:Sys}
The network environment includes multiple DCs connected to base stations. Users' equipment can send different SFC requests to the network. Each DC has VNF-instance (VNFI) capabilities, enabling the installation and allocation of VNFs to SFC requests based on its computational and storage resources (\figurename \ref{fig: env}). 
 The network topology is represented by a graph ${G}({N},{L})$, where ${N}$ denotes the set of VNFI-enabled DCs and ${L}$ represents the logical links connecting the DCs. The matrix ${W}=|{N}|X|{N}|$ quantifies the distance between DCs. Specifically, each element $w_{mn}$ in the matrix signifies the distance weight between DCs $m$ and $n$ within the network. The bandwidth (BW) capacity of the link between DCs $m$ and $n$ is represented as $B_{mn}$ Mbps.  ${\Pi}_m$ denotes the storage capacity of DC $m \in {N}$, measured in gigabytes (GB). Further, the computational capacity of DC $m \in \mathcal{N}$, measured in cycles per second, is denoted as ${\Phi}_m$ and is determined by the number of CPUs and available RAM resources in the DC.

The sets of different SFC requests supported by the service provider and of VNFs used to serve these SFCs are denoted by $\mathcal{S}$ and $\mathcal{V}$, respectively. $\mathcal{FC}^s= (v_1^s\rightarrow v_2^s \rightarrow...v_{k_s}^s)$  represents a chain of $k_s$ VNFs to be executed in sequence for successful completion of request $s\in \mathcal{S}$. Along with $\mathcal{FC}^s$, $\mathcal{B}^s \rightarrow$ BW requirement, $D^s \rightarrow$  E2E delay tolerance,  and $\Lambda_s \rightarrow$ request bundle size specify  different characteristics of each SFC $s\in \mathcal{S}$. In this work, we refer to the six most commonly referenced SFCs requests, as specified in Table \ref{tab:2}, with their defining characteristics \cite{2}. Furthermore, the storage, computational resource requirement, and processing time for VNFI $v \in \mathcal{V}$ are denoted by $\psi^v$, $\phi^v$,  and $t_p^v$ respectively.

It is assumed that multiple instances of the same or different VNFs can be installed in each DC and allocated to different SFC requests $s \in \mathcal{S}$. To process a VNF for an SFC, the same type of VNFI must be installed in a DC and allocated to it until its processing time is complete. Once done, another VNF of the same type from a different SFC can use the VNFI, or it can utilize another available instance of its type.  $\eta_{m}^v$  is an integer variable representing the number of VNFI $v\in \mathcal{V}$ placed on DC $m \in {N}$.  Efficient VNF placement and allocation to SFC requests are critical for optimizing network performance in terms of acceptance ratio and resource consumption while meeting SFC requests' E2E delay tolerance.  Let $A_s$ represent the number of SFC requests served of type $s \in \mathcal{S}$. The acceptance ratio, $\mathcal{A}_{R} = \frac{\sum_{s \in \mathcal{S}} A_s}{\sum_{s \in \mathcal{S}} \Lambda_s}$, is defined as the ratio of the total number of SFC requests served within their E2E delay limit to the total number of generated requests from all types of request bundles. The matheatical formulation for acceptance ratio maximization of SFC requests is as follows: 
\begin{figure*} 
    \centering
    \includegraphics[width=0.95\linewidth]{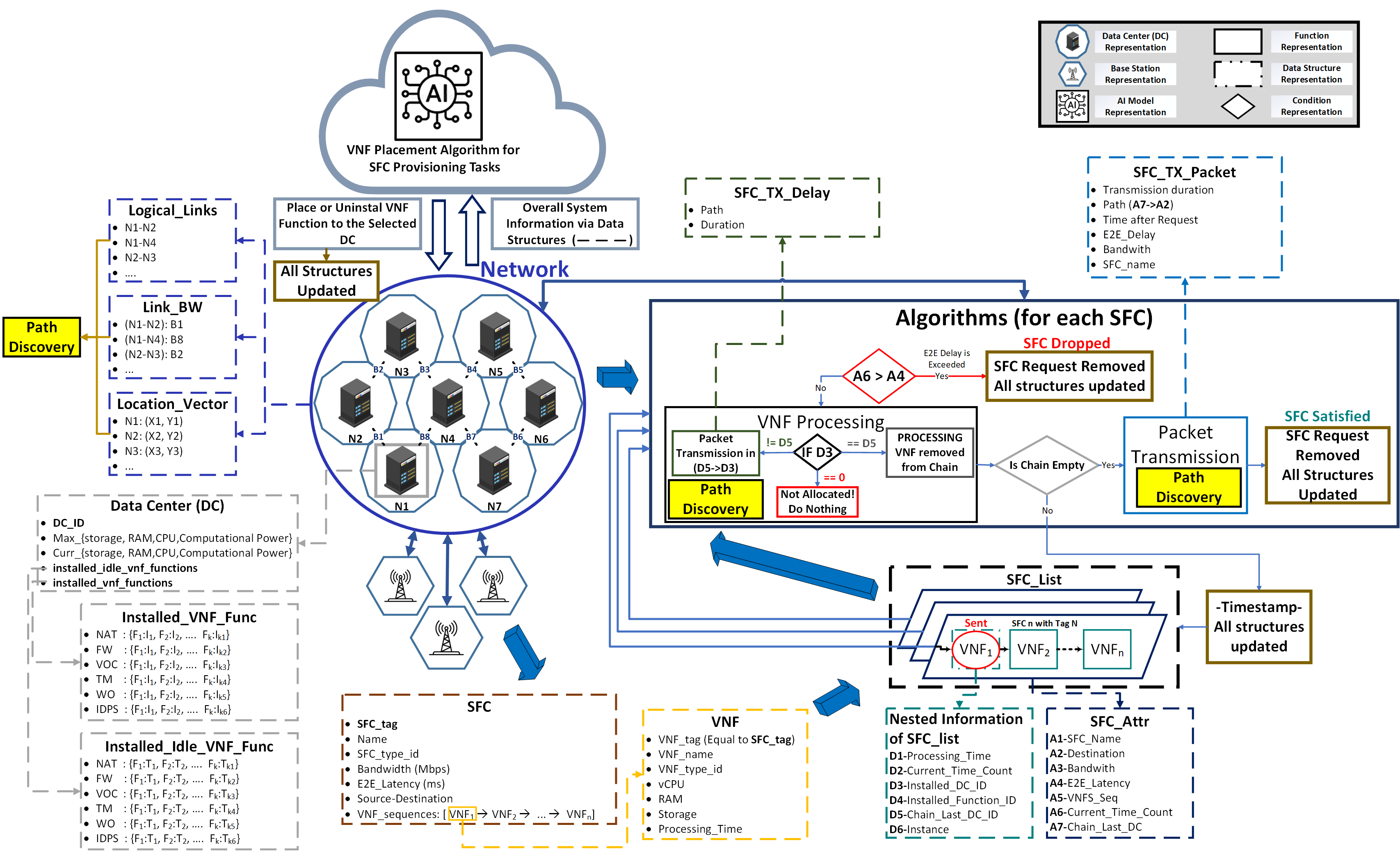}
    \caption{Platform Illustration, Data Structures, SFC Demand Creation, and AI Model for SFC Provisioning Task \vspace{-0.2in}}
    \label{fig: env}
\end{figure*}
 \begin{align*}
 &(\mathcal{P}):\;\underset{\mathbf{\eta},\boldsymbol{\Delta}}{\text{maximize}}\; \mathcal{A}_{R}\\
 &\text{s.t.:}\; (C1): \sum_{v\in \mathcal{V}} \eta_{m}^v \psi^v \leq \Pi_m, \; \forall m \in {N}\\
 & (C2):  \sum_{v\in V} \eta_{m}^v \phi^v \leq \Phi_m, \; \forall m \in {N}\\
 &(C3): \sum_{{m}\in {N}} \Delta_{m}^{v_i^s} \mathds{1}(\eta_{m}^{v_i^s}>0) = 1, \; \forall s \in \mathcal{S}, \;\;\forall i={1,2,.....k_s}\\
 & (C4): \sum_{\Lambda_s \forall s\in \mathcal{S}}\sum_{i=1}^{k_s-1} \Delta_{m}^{v_i^s} \Delta_{n}^{v_{i+1}^s} \mathcal{B}^s \leq B_{mn}, \; \forall m,n \in {N},  \;\; m\neq n  \\
 & (C5):   t_{c}^s+  t_{P_r}^s \leq {D}^s \;; \forall s\in \mathcal{S} &&
\end{align*}
Constraints $(C1)$ and $(C2)$ state that the storage and computational needs of the deployed VNFIs must not exceed the overall respective capacities of the DCs. $(C3)$ ensures that each VNF in SFC $s$ should be processed by only one DC; where, $\Delta_{m}^{v_i^s}$ is a binary variable which is set equal to $1$ if $i^{th}$ VNF in SFC $s$ is allocated to node $m$ with condition $(\eta_{m}^{v_i^s}>0)$; otherwise, it is set to zero. $(C4)$ is related to the logical link BW constraint and states that the BW resource occupied by SFCs should not exceed the BW capacity of links. Lastly, $(C5)$ confirms that communication (Propagation+TX) and processing delay of SFC should not exceed its E2E delay tolerance limit. Communication delay, $ t_{c}^s= \sum_{i=1}^{k_s-1}  \sum_{m\in {N}}  \sum_{n\in {N}} \Delta_m^{v_i^s} \Delta_n^{v_{i+1}^s} (t_{mn}^{P}+ t_{mn}^{T_X})$, where propagation delay $t_{mn}^{P}$ depends upon the distance between VNF serving nodes in SFC request $s$ and the speed of light as normally the optical fiber is used to provide a logical link among DCs i.e. $t_{mn}^P=w_{mn}/c$ with $c$ being the speed of light; and TX delay, $t_{mn}^{T_X}$ depends upon packet length and Link BW assigned to SFC request. The  processing delay, $t_{P_r}^s=\sum_{i=1}^{k_s}  \sum_{m\in {N}} \Delta_{m}^{v_i^s} (w^{v_i^s}+t_p^{v_i^s})$, depends upon the waiting time $w^{v_i^s}$ before allocation, and processing time $t_p^{v_i^s}$ of different VNFs to be executed in the SFC request $s\in \mathcal{S}$. The optimization problem $\mathcal{P}$ is combinatorial, non-convex and NP-hard. To solve this problem, an advanced DRL-based SFC provisioning algorithm and a novel realistic simulation platform are investigated.

\section{Simulation Platform and Structures} \label{sec:2}

\subsection{SFC and VNF Structures} \label{sfc}
SFC requests are generated throughout the simulation with various numbers called bundle sizes. They have the attributes source and destination DC ID, $SFC_{Name}$, Bandwidth (Mbps), and E2E latency (ms). Each SFC includes a unique ID number, which is called \textbf{SFC\_Tag}, and this tag is shared with all VNFs in the chain. In addition, SFC objects include \textbf{Type\_id}, and all features are assigned by considering it. When the simulation creates the requests, the SFC object is first created. Then, according to SFC type\_id, VNF instances are created and chained in proper order in the data structure \textbf{VNF\_Sequences} of the SFC objects. The types of SFCs used in this work are outlined in Table \ref{tab:2} with their characteristics tuple.

\begin{table} [!h]
\centering
\caption{Service Function Chain (SFC) characteristics \cite{2}}\fontsize{6.5}{7.7}\selectfont
\begin{tabular}{|p{1.63cm}|p{1.6cm}|p{0.9cm}|p{1.23cm}| p{0.9cm}|} 
 \hline
 \textbf{SFC Request}&\textbf{VNF Sequence} &\textbf{Bandwith (Mbps)} &\textbf{E2E delay (ms)} &\textbf{Request Bundle} \\  
 \hline
  Cloud \break Gaming (CG) & NAT-FW-VOC\break-WO-IDPS  & 4 & 80  & [40-55] \\
  \hline
  Augmented \break Reality (AugR) & NAT-FW-TM\break-VOC-IDPS & 100 & 10  & [1-4] \\
  \hline
  VoIP & NAT-FW-TM\break-FW-NAT & 0.064 & 100 & [100-200] \\
  \hline
  Video \break Streaming (VS) & NAT-FW-TM\break-VOC-IDPS & 4  & 100  & [50-100] \\
  \hline
  MIoT & NAT-FW-IDPS  & [1-50] & 5 & [10-15] \\
  \hline
  Ind 4.0 & NAT-FW & 70 & 8 & [1-4] \\
 \hline
\end{tabular}
\label{tab:2}
\end{table}

 The six different VNF Instance types used by these SFCs are NAT (Network Address Translator), FW (Firewalls), VOC (Video Optimization Controller), TM (Traffic Monitor), WO (WAN Optimizer), and IDPS (Intrusion Detection and Prevention System). Each VNF has $VNF_{Name}$, vCPU, RAM (GB), storage, and processing time (0.01 ms) attributes (see in \figurename \ref{fig: env}). VNF attribute values are taken from \cite{2}. VNF object also has \textbf{Type\_id}, and all attributes are assigned according to it. The multiplication of vCPU and RAM shows the required computational power resources for the VNFs. Each VNF object in the chain holds \textbf{VNF\_Tag}, which is taken from \textbf{SFC\_Tag} to discriminate its SFC in the network. In order to process VNF, a proper function with a matching type should be installed in the DC, which includes available storage (GB) and computational power. Then, VNF can be allocated to the installed functions. 

Primary data structures for SFC information in the platform are \textbf{SFC\_List}, \textbf{SFC\_Attr}, \textbf{SFC\_TX\_Delay}, and \textbf{SFC\_TX\_Packet} (see in \figurename \ref{fig: env}). All these data structures store information as key-value pairs where sfc\_tag is the key and information is the value. \textbf{SFC\_List} is used for tracking VNFs' state in the chain while \textbf{SFC\_Attr} stores specific information for the SFC chain besides VNF information. On the other hand, \textbf{SFC\_TX\_Delay}, \textbf{SFC\_TX\_Packet} are used for tracking the TX during and after VNF processing, respectively.

\subsection{Connection Structures and Shortest Path Discovery} \label{link}                        
All connections and DCs information with their position resides in \textbf{Network} object. For connections and Nodes (DC), three data structures are used, which are  \textbf{Logical\_Links}, \textbf{Link\_BW} and \textbf{Location\_Vector} (see \figurename \ref{fig: env}); \textbf{Logical\_Links} is a list of data structure lists where each inner list contains two elements: neighbor DCs' IDs, and the available bandwidth (\textbf{Link\_BW}), which is initially set to 500 Mbps, between these DCs. Upon a new packet TX, these values change according to the SFC's required BW. \textbf{Location\_Vector}  stores the X and Y coordinates of the nodes in the network alongside DC ID.  

During the packet TX phase, these three data structures are used to discover the shortest available path between source and destination. \textit{Logical\_Links} provides the extraction of neighbor nodes and the path is found by traversing through these nodes. During this process, the path length is calculated via \textit{Location\_Vector} values while BW availability is checked via \textbf{Link\_BW}. For certain shortest-path discovery, all paths between DCs with required BW availability need to be known.  To reduce processing time when searching for new paths, if the current path length exceeds that of previously found paths, the function terminates process and disregards that path. The search for alternative paths continues until all available options have been explored. 
Lengths of paths are calculated based on the distances of DCs, and the shortest path is selected. Packet TX commences on the discovered optimum path where the available BW is reduced according to the $Bandwith$ requirement of the SFC request. 


\subsection{Data Center Structures}
DCs are used for the installation and allocation of the VNF functions in the SFC chain, which is essential for satisfying incoming requests. \textbf{Location\_Vector}, which is mentioned in \ref{link}, stores the DC ID alongside the 2-dimensional Cartesian coordinates of this DC. DC object is constructed on this location, and each DC has unique \textbf{DC\_ID}. DC includes max\_\{CPU, RAM, Storage, ComputationalPower\}, Current\_\{CPU, RAM, Storage, ComputationalPower\} attributes, and \textbf{Installed\_VNF\_Func} and \textbf{Installed\_Idle\_VNF\_Func} structures which can be seen in \figurename \ref{fig: env}. 
During the initialization of the DC object, Current attributes are set to their maximum values. 

Installation and allocation of the VNF functions are tracked in \textbf{Installed\_VNF\_Func} structure. This structure includes six different nested structures with a key of VNF type name. Each nested dictionary holds \{$\mathcal{F}_{id}$, $\mathcal{I}_{id}$\} pair where $\mathcal{F}_{id}$ is the installed function id, unique for each nested structure, and $\mathcal{I}_{id}$ is the binary value indicates status, either in use (1), allocated to any VNF with matching type, or idle (0). \textbf{Installed\_Idle\_VNF\_Func} structure, on the other hand, is used for automatically uninstallation of idle VNF function which waiting time reaches threshold limit $\mathcal{T}_{thresh}$  for the aim of minimizing the resource and energy consumption. Its architecture is very similar to \textit{Installed\_VNF\_Func} structure, but instead of status value $\mathcal{I}_{id}$, it stores idle waiting time $\mathcal{T}_{id}$ of the function. 



DC object includes various methods that allow Installation, Uninstallation, Allocation, and revoke. For error checking, an action assertion is performed, such as checking the resource availability in the installation or the function's existence in uninstallation. \textbf{Install\_VNF} takes VNF types and installs the VNF function as long as resources are available. It generates new $F_{id}$ for function and updates the \textit{Installed\_VNF\_Func} structure. The same function is also stored initially in \textit{Installed\_Idle\_VNF\_Functions} structure since it is not allocated in the beginning.  \textbf{UnInstall\_VNF} function takes VNF type and $\mathcal{F}_{id}$ and remove \{$\mathcal{F}_{id}$, $\mathcal{I}_{id}$\} pair from the structures then restores the resource. \textbf{Allocation\_VNF} takes VNF and $\mathcal{F}_{id}$ as parameters and allocate it to the appropriate SFC. This function's id is removed from the \textit{Installed\_Idle\_VNF\_Func}. Installation, allocation, and uninstallation are performed using the AI model for this research. \textbf{Revoke\_VNF} function, on the other hand, is automatically called once the allocated function's processing is completed. It changes the state of this function and transfers it into \textit{Installed\_Idle\_VNF\_Func}. However, this function requires completion of the VNF processing; if E2E delay exceeds and there are pending VNFs in the chain \textbf{Force\_Revoke\_VNF} is called.


\subsection{SFC Request Generation}
The number of SFC requests is determined by its bundle size, and the bundle size of each type of SFC request is randomly generated within a specified range as given in Table \ref{tab:2}. 
The source and destination of each SFC request are randomly or manually assigned, and every SFC has a unique tag (SFC\_tag). When a request is created, an SFC object and its corresponding VNF instances are generated and sequenced. The shared SFC tag enables easy association between VNFs and their corresponding SFC requests. This association is crucial for the VNF placement algorithm, which relies on SFC tags to determine available VNF instances and extract SFC chain information from data structures in Section \ref{sfc}. 
Once SFC generation is complete, these data structures, which are \textit{SFC\_List}, \textit{SFC\_Attr}, \textit{SFC\_TX\_Delay}, \textit{SFC\_TX\_Packet}, are updated by inserting new \{key, value\} pair where key is the SFC\_tag and value is the default initial set. For \textit{SFC\_List}, initial values are nested dictionaries consisting of its VNF states. Initial values of these nested lists are \{$\mathcal{T}_{req}$, -1, 0, 0,  $DC_{srcId}$, $VNF_{ins}$]. 
These key-value pairs are either changed or removed from the data structures according to the actions and simulation runtime.

\textbf{SFC\_List} is one of the primary dictionary-type data structures in the simulation that takes \textit{SFC\_Tag} as key and VNFs information in the chain as value. Value of the  SFC\_List is also a nested dictionary which takes $VNF_{Name}$ as key and VNF state as the value for each VNF in the chain. The value of this nested dictionary is a state list that consists of \{$\mathcal{T}_{req}$, $\mathcal{T}_{Vcurr}$, $VNF_{DCId}$, $VNF_{funcId}$, $SFC_{DCId}$, $VNF_{ins}$\}. $\mathcal{T}_{req}$ is equal to the VNF object's processing time duration, which is the static value of the state. $\mathcal{T}_{Vcurr}$ shows the time after VNF is allocated. Initially, this state starts with -1 and stays the same until VNF allocation, which turns this value into 0 and starts counting. $VNF_{DCId}$ shows which DC this VNF is allocated to and starts with 0, which means the VNF is not allocated yet. $VNF_{funcId}$ indicates the function ID used for allocation, and it starts with 0. $SFC_{DCId}$ shows the DC ID, which is the SFC Chain's last position. Initially, it starts with $DC_{srcId}$ since SFC requests' processing starts in the source DC. Finally, $VNF_{ins}$ is the VNF object in the chain.

 \textbf{SFC\_Attr} stores the SFCs' attributes, beside their VNFs sequence state,  which includes \{$SFC_{Name}$, $DC_{destId}$, $SFC_{BW}$, $\mathcal{T}_{E2E}$, $VNF_{Seq}$, $\mathcal{T}_{Ccurr}$, $SFC_{DCId}$\}. $SFC_{Name}$, $DC_{destId}$, $SFC_{BW}$, $\mathcal{T}_{E2E}$ and $VNF_{Seq}$ are static values and placed once SFC is created. $SFC_{DCId}$ denotes the current data center ID for an SFC request, initially mirroring the request source. Once VNF functions are allocated, it switches to the respective DC ID.  $\mathcal{T}_{Ccurr}$ tracks the time passed since request initiation, starting at 0. For \textbf{SFC\_TX\_Delay}, \textbf{SFC\_TX\_Packet}, they store information as \{$Path_{VNF}$, $\mathcal{T}_{Remain}$\} and \{$\mathcal{T}_{Remain}$, $Path_{SFC}$, $\mathcal{T}_{Ccurr}$, $\mathcal{T}_{E2E}$, $SFC_{BW}$, $SFC_{Name}$\}.

\subsection{Simulation Run and VNF Processing}



The simulation model runs from SFC request generation until all SFC requests in $SFC\_List$ are either processed or dropped on $\mathcal{T}_{Ccurr}$ exceeding their $\mathcal{T}_{E2E}$. \textit{SFC\_List} holds each VNF state information of the SFC requests. Once one VNF is allocated, its $\mathcal{T}_{Vcurr}$ value turns from -1 to 0, which indicates that it can start counting in every timestamp. Moreover, $VNF_{DCId}$ and $VNF_{funcId}$ of this VNF state turn into the DC and function ID in which it is installed. However, If $VNF_{DCId}$ and $SFC_{DCId}$ values are not the same, it requires data TX from $SFC_{DC}$ to $VNF_{DC}$ during the VNF processing phase, which requires the updating of \textit{SFC\_TX\_Delay}. After packet TX is completed, $SFC_{DCId}$ takes the value of $VNF_{DCId}$, and processing of the current VNF starts. In every timestamp, $\mathcal{T}_{Vcurr}$ increases to 1, which is also called \textbf{step} of the simulation. Once the $\mathcal{T}_{Vcurr}$ value reaches the $\mathcal{T}_{req}$, it means VNF processing is completed. First, VNF is automatically revoked from its DC by looking up $VNF_{DCId}$ and $VNF_{funcId}$  values. Then, this VNF state is removed from the chain in \textit{SFC\_List} with \textit{SFC\_Tag}. If this chain in \textit{SFC\_List} is empty, it indicates that all VNF functions are completed, and the VNF processing phase for this request is done. Afterward, \textit{SFC\_TX\_Packet} values related to the SFC are updated, and the rest of the data structures removed this SFC information from their storage. Finally, the packet TX phase starts, and SFC requests are satisfied after the packets reach the DC with $DC_{destId}$ ID. 

In the platform, it is crucial to monitor unique cases. For example, $\mathcal{T}_{Ccurr}$ value in \textit{SFC\_Attr} is checked with $\mathcal{T}_{E2E}$ and if this value reaches the $\mathcal{T}_{E2E}$, SFC requests drop automatically, and 
if there are any assigned functions, they are instantly revoked. Further, if the route for any packet TX phase is not discovered, this function must wait until the next step. It is crucial to recognize the need for VNF functions to execute in a specific sequence within the chain. This means a VNF cannot start processing until the preceding higher-order function in the SFC finishes. Algorithm 1 details the SFC provisioning phase via VNF placement, as depicted in \figurename \ref{fig: env}.

In algorithm 1, \textbf{items} in the \textit{SFC\_List} denote the nested values for each VNF information which are \{$\mathcal{T}_{req}$, $\mathcal{T}_{Vcurr}$, $VNF_{DCId}$, $VNF_{funcId}$, $SFC_{DCId}$, $VNF_{ins}$\} with starting index 0. Before processing, it first checks the E2E latency; then, it iterates all SFC requests by looking up only the first VNFs in the chain. In the final loop, if an empty SFC chain is found, packet TX for this request starts. 

\section{SFC Provisioning Algorithm} \label{sec:3}



\subsection{Deep Reinforcement Learning Module}

The DRL architecture employs fully connected DNN layers to process environmental information, using advanced architectures including multiple input layers for each feature and attention layer, encompassing DC and SFC states. These layers determine the action model's output. Decision-making in this setup is represented using a Markov Decision Process (MDP) with a DQN algorithm, which solves MDPs by approximating the optimal action-value function known as the Q-value. MDP comprises states, actions, rewards, and next states, and the primary aim is to establish a policy that maximizes the cumulative expected reward over time.  In RL, the \textbf{State} encompasses the information available to an agent regarding the environment at any specific moment. This information is propagated through the hidden layers of the model, guiding the selection of actions aimed at maximizing rewards. In the proposed model, State values take the value of the SFC and DC information, which are collected from \ref{sec:2}.

\SetAlgoSkip{}
\begin{algorithm} [!h]

\caption{Simulation Step for VNF Processing}
\SetAlgoLined
\SetAlgoNlRelativeSize{-2} \fontsize{9}{9}\selectfont 
\KwData{SFC\_List, SFC\_TX\_Delay, SFC\_Attr}
\KwResult{Process VNFs in SFCs}

\SetKwFunction{getPacketLen}{getPacketLen}
\SetKwFunction{setLastDC}{setLastDC}
\SetKwFunction{getSFCsTagList}{getSFCsTagList}
\SetKwFunction{StartPacketTX}{StartPacketTX}
\SetKwFunction{UpdatePacketTXStructure}{UpdatePacketTXStructure}
\SetKwFunction{getVNFsInsSFCChain}{getVNFsInsSFCChain}

\fontsize{7.0}{7.0}\selectfont 

$list\_of\_tag \leftarrow \getSFCsTagList()$

\ForEach{$tag$ \textbf{in} $list\_of\_tag$}{
    \If{$SFC\_Attr[tag][``\mathcal{T}_{Ccurr}"] > 100 \times SFC\_Attr[tag][``\mathcal{T}_{E2E}"]$ \textbf{and} $SFC\_List[tag] \neq \{\}$}{
        $VNF\_list \leftarrow getVNFsInsSFCChain(SFC\_List,tag)$ 
        
        \ForEach{$VNF$ \textbf{in} $VNF\_list$}{
            \If{$SFC\_List[tag][VNF][1] \neq -1$}{
                $force\_Revoke\_VNF(SFC\_List[tag][VNF][2], \\\phantom{for} SFC\_List[tag][VNF][5], SFC\_List[tag][VNF][3])$
                
                \textbf{delete} $SFC\_List[tag][VNF]$
            }
        }
        
        \textbf{delete} $SFC\_\{list,Attr,TX\_delay,TX\_Packet\}[tag]$
        
        \algorithmiccomment{\textbf{SFC Dropped}}\;
    }
}

\ForEach{$tag, sub\_SFC$ \textbf{in} $SFC\_List.items()$}{
    \ForEach{$VNF, items$ \textbf{in} $sub\_SFC.items()$}{
        \If{$SFC\_TX\_Delay[tag][1] \neq 0$ \textbf{and}  $SFC\_TX\_Delay[tag][0] \neq []$}{
            $SFC\_TX\_Delay[tag][1] \leftarrow SFC\_TX\_Delay[tag][1] - 1$
            
            \If{$SFC\_TX\_Delay[tag][1] \leq 0.0$}{$LA\_Network.update\_bw(SFC\_TX\_Delay[tag][0],\\\phantom{forforforforforforfor}SFC\_Attr[tag][``BW"])$
            
                $SFC\_TX\_Delay[tag][0] \leftarrow []$\;
            }
        }
        \ElseIf{$items[1] \neq -1$ \textbf{and} $items[0] - 1 \neq items[1]$}{
            \If{$items[4] == 0$}{
            
                
                $curr\_dc \leftarrow items[2]$ \textbf{(No VNFs in Chain Start Processing)}
                
                $\setLastDC(VNF_{DC}, VNF\_in\_Chain)$
                
                $SFC\_Attr[tag][``SFC_{DC}"] \leftarrow VNF_{DC}$\;
            }
            \ElseIf{$items[4] == items[2]$}{ 
                
                $items[1] \leftarrow items[1] + 1$\; \textbf{(VNF Processing)};
                
            }
            \ElseIf{$items[4] \neq items[2]$}{
            
                
                $cur\_dc \leftarrow items[2]$ \textbf{(VNF's SFC is different then SFC's DC)}
                
                $length, path \leftarrow LA\_Network.select\_min\_path(\\\phantom{forfor} src=items[4], dest=items[2], \\\phantom{forfor} req\_bw=SFC\_Attr[tag][``BW"])$
                
                \If{$path \neq []$}{
                
                    \algorithmiccomment{\textbf{Packet TX During VNF Processing Phase}}
                    
                    \setLastDC(curr\_dc, VNF\_in\_Chain)
                    
                    $packet\_len \leftarrow getPacketLen(SFC\_Attr[tag])$ \textbf{Mb}
                    
                    $TX\_Time \leftarrow$ $packet\_len / SFC\_Attr[tag][``BW"]$
                    
                    $SFC\_TX\_Delay[tag] \leftarrow [path, Trans\_Time]$
                    
                    $Network.update\_bw( path, SFC\_Attr[tag][``BW"])$
                    
                    $SFC\_Attr[tag][``SFC_{DC}"] \leftarrow curr\_dc$\;
                }
            \textbf{else} \textbf{then} \{ No Path For TX; \}
            }
        }
        \ElseIf{$items[0] - 1 == items[1]$}{
            
            $Revoke\_VNF(items[2], items[5], items[3])$
            
            \textbf{delete} $sub\_SFC[VNF]$\;
        }
        \textbf{break} \textit{(Current Step Looks Only First VNF in SFC)}\;
    }
    $SFC\_Attr[tag][``\mathcal{T}_{Ccurr}"] \leftarrow SFC\_Attr[tag][``\mathcal{T}_{Ccurr}"] + 1$\;
} 
$list\_of\_tag \leftarrow \getSFCsTagList()$

\ForEach{$tag$ \textbf{in} $list\_of\_tag$}{
    \If{$SFC\_List[tag] == \{\}$}{
    
        \UpdatePacketTXStructure(tag)
        
       \textbf{delete} $SFC\_\{List,Attr,TX\_Delay\}[tag]$
       
        \StartPacketTX(tag,$SFC\_TX\_Packet$); \algorithmiccomment{\textbf{All VNFs are Processed}}

    }
}

\end{algorithm}

\textbf{Action} is a range of possible scenarios modeled after receiving state data from the environment. The DRL model outputs the agent's actions, which include \textit{Allocate\_VNF}, \textit{Uninstall\_VNF} and \textit{Idle\_Wait}. \textit{Allocate\_VNF} assigns VNFs if any idle ones exist; otherwise, it installs a new VNF and assigns it. \textbf{Rewards} provide feedback on the agent's performance for each action, with penalties for sub-optimal choices. Processing all VNF functions in the SFC chain yields a positive reward, while dropping the SFC incurs a penalty. The algorithm operates every $\mathcal{T}_{Model}$ ms, updating the DRL model's state to the \textbf{Next State} after each action.

The output of the model is the decision of the \textbf{VNF\_Type}, and \textbf{Action\_Type} of the algorithm if the action type is \textit{Uninstall\_VNF} or \textit{Idle\_Wait}, model executes its action directly. However, if the action type is \textit{Allocate\_VNF}, it requires \textbf{Priority Points} of the VNFs.
\vspace{-0.1in}
\subsection{Priority Points}
Upon selecting the action as placing, the model picks the most fitting VNF functions from the pool awaiting allocation. To assess their significance, it gathers selected function types and assigns priority points based on criteria such as previous VNF allocation in the same DC or minimum remaining E2E latency time for the SFC request. This process involves evaluating four distinct criteria.

The first priority is the remaining time of VNF functions before their SFC chain drops. 
The second priority shows the importance of the DC for VNF, whether it's the source (max priority), on the path to the destination, or none (zero priority) at all.
The third priority is to consider whether any function of the VNF's SFC is deployed in the current DC. If so, it is positive; otherwise, zero. The fourth priority is the urgency of VNF functions, raising priority if time left is less than the threshold $\mathcal{T}_{urgency}$. Finally, the function with the highest combined priority is selected for placement.  

\section{Numerical Results} \label{sec:4}

\begin{figure} [!h]
    \centering
    \includegraphics[width=1.0\linewidth]{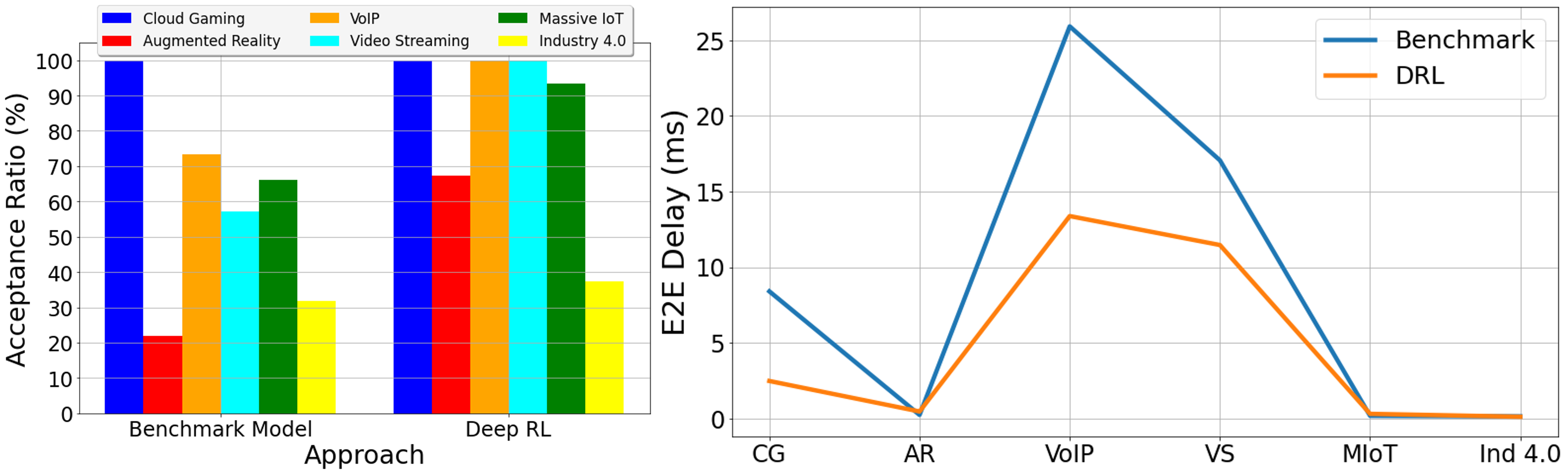}
    \caption{SFC acceptance ratio and E2E delay for each type of requests with 5 DCs}
    \label{fig: res 1}
     \vspace{-0.1in}
\end{figure}

We configure networking settings for SFC provisioning tasks and test our AI module's performance in the proposed platform. We set up 5 data centers, each with 2000 GB storage, 64 CPUs, and 256 GB RAM. The network faces a high volume of SFC requests with various bundle size ratios (see Table \ref{tab:2}). Our platform's efficacy in SFC Provisioning is verified through performance testing using advanced models, producing multiple runtime results. Testing occurs dynamically with SFC requests generated four times in the simulation. \figurename \ref{fig: res 1}  displays the E2E delay and acceptance ratio comparison between our DRL model and the benchmark model, which relies solely on a heuristic approach, for each SFC request type.  

\begin{figure} [!h]
    \centering
    \includegraphics[width=1.0\linewidth]{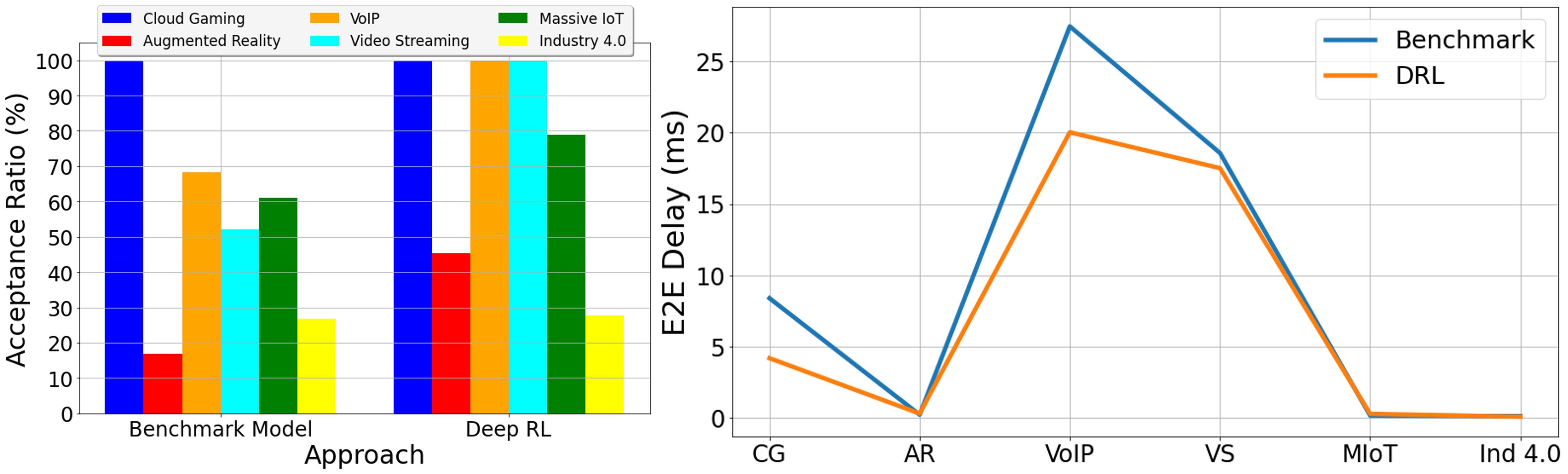}
    \caption{SFC acceptance ratio and E2E delay for each type of requests with 3 DCs}
    \label{fig: res 1.5}
    \vspace{-0.1in}
\end{figure}

According to \figurename \ref{fig: res 1}, for the benchmark models, AugR and Ind4.0 have the minimum acceptance ratio, which is 20\% and 30\%. Also, MIoT, VS, and VoIP's acceptance ratios are around 65\%. On the other hand, the Deep RL model can satisfy all CG, VoIP, and VS requests while the acceptance ratio of AugR and Ind4.0 increases to around 68\% and 48\%. Since E2E delays are calculated using only accepted SFC requests, AugR's delay in the benchmark model is slightly lower than DRL's. However, the E2E delays of CG, VoIP, and VS have reduced significantly. Because MIoT and Ind4.0's E2E delay tolerances are so low, such as $5$ms and $8$ms, these SFC requests are served at a higher priority, so their latencies are minimum. Resource availability is a significant factor in SFC acceptance, therefore to show the results under different resources and reconfigurability for our proposed approach, network configuration is changed, and the number of DC is reduced to 3. New results can be seen in \figurename \ref{fig: res 1.5}.

\begin{figure} [!h]
    \centering
    \includegraphics[width=1\linewidth]{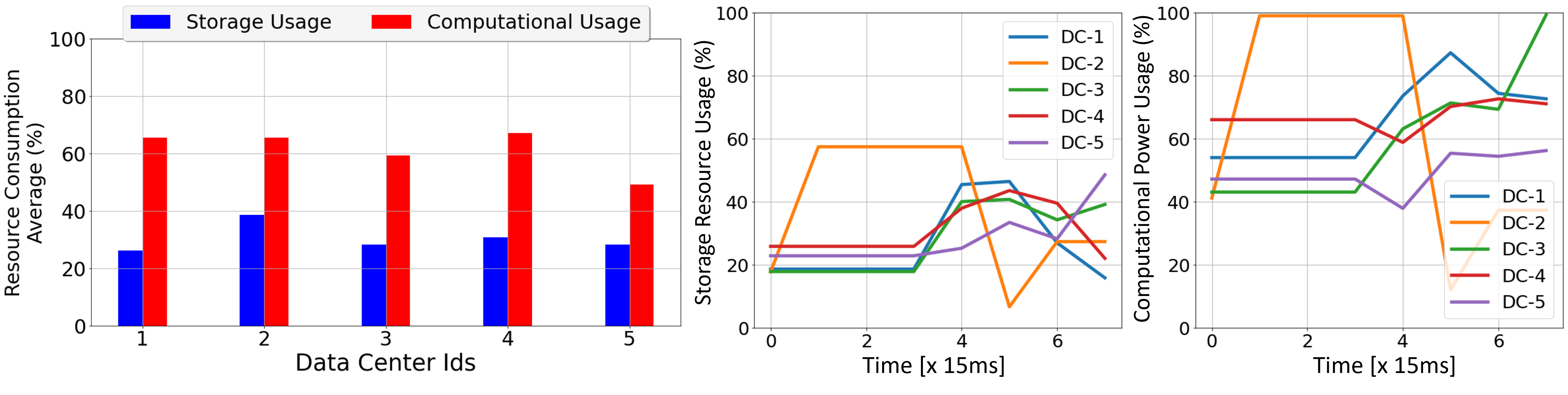}
    \caption{Resource consumption: average  and during simulation at 15ms time scale }
    \label{fig: res 2}
     \vspace{-0.1in}
\end{figure}

In \figurename \ref{fig: res 1.5}, once resources are minimized, the ratios of accepted SFCs in both algorithms are reduced. Although the DRL model preserves the acceptance ratio of CG, VoIP, and VS type SFC requests, the acceptance ratio for AugR, MIoT, and Ind4.0 decreased, mostly in AugR, around 15\%. Moreover, in the benchmark approach, except for CG, the SFC acceptance ratio decreases for every SFC type. On the other hand, tracking resource consumption is a major factor in inferring acceptance ratio results. Resource consumption in terms of storage and computational power for each DC, for a network with 5 DCs, can be seen in \figurename \ref{fig: res 2}.

In \figurename \ref{fig: res 2}, the left part shows average resource consumption during simulation runtime, while the right part illustrates resource usage for each timestamp in the run, which is essential to track DC situations in a real-time manner. To preserve memory in testing, results are collected every 15 ms of a simulation run. 
For average, storage is used around 25\% while for computational power, it is around 62\%. Resource consumption of DC indicates VNF processing cost, hence ideal VNF should be uninstalled to reduce resource consumption.  According to the right part results, DC\_2 is used at almost full capacity till the middle of simulation. Then, its resource consumption is reduced by uninstallation of idle VNFs.  Other DCs' usage increases depending on new SFC request generation during simulation run. 

\section{Conclusions} \label{sec:5}
This paper has introduced a new practical platform suitable for various studies on SFC provisioning tasks. It takes into account various types of VNFs and SFCs, each characterized by unique attributes, including chain sequence, storage and computational requirements, E2E delay, and more. Also, platform supports execution of diverse AI models and methodologies, enabling the collection of different results such as E2E latency, acceptance ratio, and resource consumption for further analysis. We have also introduced a DRL model incorporating priority points for SFC provisioning tasks aimed at meeting demands via an effective VNF placement algorithm. In this algorithm, DRL module receives network-related inputs to make decisions regarding actions, while priority points are allocated to VNFs to enhance robustness of performance. Both network and DRL models exhibit reconfigurable behavior, with E2E delay, acceptance ratio, and resource consumption serving as metrics to assess performance across various configurations. Results have shown that DRL model enhances acceptance ratio and reduces E2E delay. Furthermore, resource consumption is monitored to infer deeper insights. 
 \vspace{-0.08in}

\section*{Acknowledgment }\label{sec:6}
This work is supported by the Natural Sciences and Engineering Research Council of Canada (NSERC) Alliance Program, MITACS Accelerate Program, and NSERC CREATE TRAVERSAL program.

\bibliographystyle{IEEEtran}

\end{document}